\documentclass[twocolumn]{aa}
\usepackage{graphicx}
\usepackage{txfonts}

\begin{document}

   \title{Dwarf elliptical galaxies with kinematically decoupled cores}

%   \subtitle{Possible evidence for galaxy harassment}

\author{S. De Rijcke \inst{1}\fnmsep\thanks{Postdoctoral Fellow of the
Fund for Scientific Research - Flanders (Belgium)(F.W.O)}, H. Dejonghe
\inst{1}, W.~W. Zeilinger \inst{2} \and G.~K.~T. Hau \inst{3} }

\offprints{S. De Rijcke}

\institute{Sterrenkundig Observatorium, Ghent University, Krijgslaan
281, S9, B-9000 Gent, Belgium \\ \email{sven.derijcke@UGent.be}, \\
\email{herwig.dejonghe@UGent.be} \and Institut f\"ur Astronomie,
Universit\"at Wien, T\"urkenschanzstra{\ss}e 17, A-1180 Wien, Austria
\\ \email{zeilinger@astro.univie.ac.at} \and ESO,
Karl-Schwarzschild-Strasse 2, D-85748 Garching bei M\"unchen, Germany
\\ \email{ghau@eso.org} }

\date{Received September 15, 1996; accepted March 16, 1997}

\abstract{We present, for the first time, photometric and kinematical
evidence, obtained with FORS2 on the VLT, for the existence of
kinematically decoupled cores (KDCs) in two dwarf elliptical galaxies;
FS76 in the NGC5044 group and FS373 in the NGC3258 group. Both
kinematically peculiar subcomponents rotate in the same sense as the
main body of their host galaxy but betray their presence by a
pronounced bump in the rotation velocity profiles at a radius of about
1$''$. The KDC in FS76 rotates at $10 \pm 3$~km/s, with the host
galaxy rotating at $15 \pm 6$~km/s; the KDC in FS373 has a rotation
velocity of $6 \pm 2$~km/s while the galaxy itself rotates at $20 \pm
5$~km/s. FS373 has a very complex rotation velocity profile with the
velocity changing sign at 1.5~$R_{\rm e}$. The velocity and velocity
dispersion profiles of FS76 are asymmetric at larger radii.  This
could be caused by a past gravitational interaction with the giant
elliptical NGC5044, which is at a projected distance of 50~kpc. We
argue that these decoupled cores are most likely not produced by
mergers in a group or cluster environment because of the prohibitively
large relative velocities. A plausible alternative is offered by flyby
interactions between a dwarf elliptical or its disky progenitor and a
massive galaxy. The tidal forces during an interaction at the relative
velocities and impact parameters typical for a group environment exert
a torque on the dwarf galaxy that, according to analytical estimates,
transfers enough angular momentum to its stellar envelope to explain
the observed peculiar kinematics.   }

\titlerunning{dEs with KDCs}

   \maketitle
%
%________________________________________________________________

\section{Introduction}

Dwarf elliptical galaxies (dEs) are small, low-luminosity galaxies
with diffuse, exponentially declining surface-brightness profiles
(\cite{fb}). They are a gregarious species and are found abundantly in
clusters and groups of galaxies (although they seem to avoid the very
cluster center where the tidal forces exerted by the cluster potential
are strong enough to disrupt them (\cite{tru})).  According to one
model for dE evolution, they are primordial objects.
% whose dynamics are dominated by a massive dark matter halo. 
Supernova explosions heat the interstellar gas to temperatures
exceeding the escape velocity, expelling gas from the galaxy
(\cite{ds,my}). This scenario explains the diffuse appearance of dEs
with enhanced star formation at larger radii. They are expected to
form a homogeneous class and to have properties that correlate tightly
with mass. Alternatively, dEs could stem from late-type disk galaxies
that entered the clusters and groups of galaxies about 5~Gyr ago
(\cite{co}). $N$-body simulations show that high-speed gravitational
interactions trigger bar-formation in any small disk galaxy orbiting
in a cluster (\cite{mkldo}) or around a massive galaxy in a group
environment (\cite{ma}) and strip large amounts of stars, gas, and
dark matter from it by tidal forces. Internal dynamical processes
subsequently transform a disk galaxy into a dynamically hot spheriodal
dE within a timespan of about 5~Gyr. 
%Since large amounts of dark matter have been stripped away, the 
%simulations predict that dEs still possess dark-matter halos albeit 
%with low mass-to-light ratios. 
Some dEs might still contain a memory of their former state. 
%Such
%signatures have indeed been observed in a rising number of
%dEs. 
Examples are dEs with embedded stellar disks, bars, and spiral
structure (\cite{ba,dr2,gr03}) and with sizable amounts of warm gas,
suggesting recent star formation in some dEs
(\cite{dr3,dm}). Moreover, rotationally flattened dEs have been
discovered (\cite{dr1,sp}). The harassment model also offers a natural
explanation for the Butcher-Oemler effect (\cite{bo}) and the
morphology-density relation (\cite{ma}).
%, and the presence of an
%intra-cluster medium as the debris of harassed galaxies
%(\cite{mkldo}).

\begin{figure}
\vspace*{6.9cm}
\special{hscale=65 vscale=65 hsize=250 vsize=190
hoffset=-40 voffset=-260 angle=0 psfile="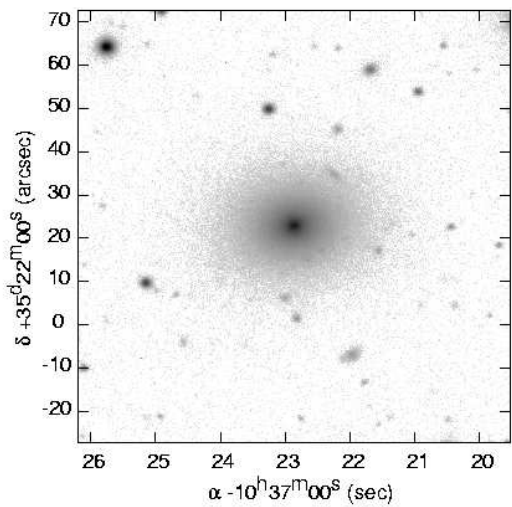"}
\caption{270~sec. I-band image of FS373, a dE2,N in the NGC3258
group. North is up, east is left. \label{ima373}}
\end{figure}
\begin{figure}
\vspace*{6.75cm}
\special{hscale=65 vscale=65 hsize=700 vsize=190
hoffset=-40 voffset=-260 angle=0 psfile="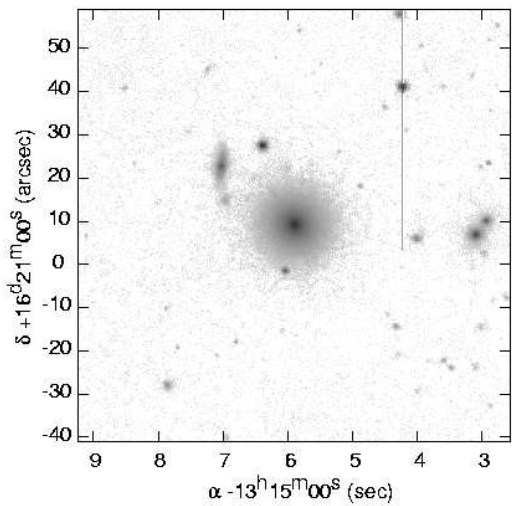"}
\caption{300~sec. I-band image of FS76, a dE0,N in the NGC5044
group. North is up, east is left. \label{ima76}}
\end{figure}

In this paper, we present photometric and kinematical evidence for the
presence of kinematically decoupled cores (KDCs) in two dEs in a group
environment:~FS373 and FS76 (we use the galaxy identification numbers
introduced by \cite{fe90}). FS373 (Fig. \ref{ima373}) is a nucleated
dwarf elliptical (dE2,N) in the NGC3258 group at a distance of 41~Mpc
(we use $H_0=70$~km/s/Mpc throughout the paper). FS76
(Fig. \ref{ima76}) is a dE0 in the NGC5044 group, at a distance of
36~Mpc.The pronounced bump in the rotation velocity profiles signals
the presence of a dynamically peculiar component in corotation with
the main body of these galaxies. Both in FS76 and FS373, the KDC
dominates the kinematics out to a radius of $~1.5''-2''$, which is
well outside the nucleus or the central density cusp. Hence, the KDC
should not be associated with the nucleus in the center of the host
dE. It is the first time that evidence is found for the existence
of KDCs in dwarf elliptical galaxies.

Massive elliptical galaxies with KDCs both in corotation and in
counterrotation with the host galaxy (e.g. \cite{ef}, \cite{ca02}) and
ellipticals with peculiar central kinematics (\cite{fiz}) have been
known for a long time. \cite{bs} found KDCs in ellipticals to be more
metalrich than the galaxies' main bodies. A merger of a giant and a
dwarf galaxy was a plausible way of producing KDCs with the observed
properties (\cite{bq}). The KDC's angular momentum vector is set
predominantly by the engulfed dwarf's angular momentum and need not be
aligned with that of the giant galaxy. KDCs in S0s can equally well be
explained by the merger of two unequal-mass spiral galaxies
(\cite{bg}). KDCs thus provide strong evidence that mergers of
(gasrich) progenitor galaxies played an important role in the past
evolution of bright elliptical and lenticular galaxies, corroborating
the hierarchical merging model for cosmological structure formation.

In section \ref{obs}, we discuss the details of the observations and
the data-reduction process. The stellar kinematics, photometry, and
measurements of the strength of the near-infrared Ca{\sc ii} triplet
(quantified by the CaT index) of these objects are presented in
section \ref{kin}, followed by a study of their internal dynamics in
section \ref{dyn}. The significance of these results in the light of
the existing theories for dE evolution is discussed in section
\ref{dis}. We summarize our conclusions in section \ref{con}.

\section{Observations and data reduction} \label{obs}

Within the framework of an ESO Large Program, we collected Bessel
VRI-band images and deep major and minor axis spectra with
unprecedented spatial and spectral resolution of a sample of 15 dEs
and dwarf lenticulars (dS0), both in group (NGC5044, NGC5898, and
NGC3258 groups) and cluster environments (Fornax cluster). The data
were taken with the FORS2 imaging spectrograph mounted on the VLT. The
images were bias-subtracted and flatfielded using skyflats taken
during twilight of the same night as the science frames. The sky
background was removed by fitting a tilted plane to regions of the
images free of stars or other objects and subtracting it. The
photometric zeropoints in each band were measured using photometric
standard stars observed during the same night as the science
frames. The images were corrected for airmass and interstellar
extinction, using the Galactic extinction estimates from
\cite{schlegel98}.

The spectra, with typical exposure times of $5-8$~h per position angle
and a seeing in the range $0.3''-1.0''$~FWHM, cover the wavelength
region around the strong Ca{\sc ii} triplet absorption lines ($\sim
8600$~{\AA}). All standard data reduction procedures
(bias-subtraction, flatfielding, cosmic removal,
wavelength-calibration, sky-subtraction, flux-calibration) were
carried out with {\tt ESO-MIDAS}\footnote{{\tt ESO-MIDAS} is developed
and maintained by the European Southern Observatory}, {\tt
IRAF}\footnote{{\tt IRAF} is distributed by the National Optical
Astronomy Observatories, which are operated by the Association of
Universities for Research in Astronomy, Inc., under cooperative
agreement with the National Science Foundation.}, and our own
software. Fitting the dispersion relation by a cubic spline, the lines
of the arc spectra are rectified to an accuracy of $1-2$~km/s FWHM.
We extracted the stellar kinematical information by fitting a weighted
mix of late G to late K giant stars, broadened with a parameterised
line-of-sight velocity distribution (LOSVD) to the galaxy spectra. We
approximated the LOSVD by a fourth-order Gauss-Hermite series
%condensing the kinematics to the mean velocity $v_p$, the velocity
%dispersion $\sigma_p$, and two coefficients $h_3$ and $h_4$,
%quantifying respectively asymmetric and symmetric deviations from a
%Gaussian LOSVD 
(\cite{gh,vf}) (the kinematics of the full sample will be presented in
a forthcoming paper). The strong Ca{\sc ii} lines, which contain most
of the kinematical information, are rather insensitive to the age and
metallicity of an old stellar population (see \cite{mi03} and
references therein), so template mismatch does not significantly
affect the results. The spectra contain useful kinematical information
out to 1.5-2 half-light radii ($R_{\rm e}$).

This is the first time a data set of dE kinematics is assembled on a
par with what so far has been achieved for bright elliptical
galaxies. Thanks to the excellent quality of the spectra, both in
terms of instrumental resolution and of seeing, we are able to
spatially resolve small-scale structures in the kinematic profiles.

\section{Kinematics, line-strengths, and photometry} \label{kin}

\begin{figure}
\vspace*{8cm} \special{hscale=90 vscale=90 hsize=700 vsize=300
hoffset=-25 voffset=-330 angle=0 psfile="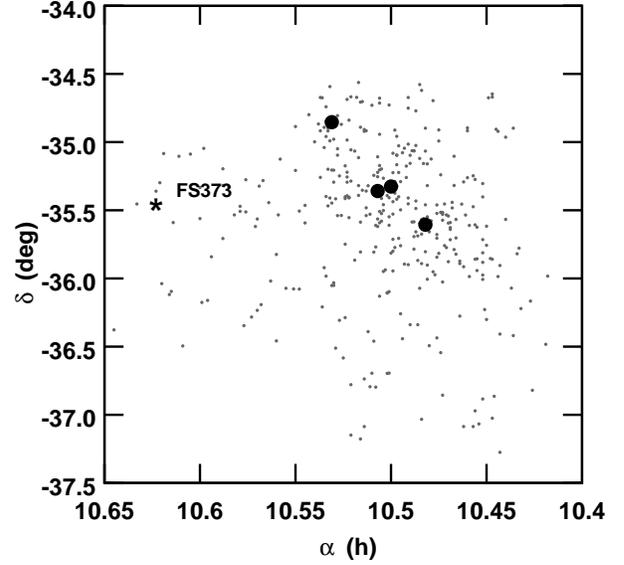"}
\caption{Position of FS373 (asterisk) inside the NGC3258 group. Small
dots indicate the positions of the 375 group members listed in
NED. Large dots indicate the positions of galaxies brighter than $M_B
= -20$. \label{catFS373ps}}
\end{figure}

\begin{figure}
\vspace*{8cm}
\special{hscale=90 vscale=90 hsize=700 vsize=300
hoffset=-25 voffset=-330 angle=0 psfile="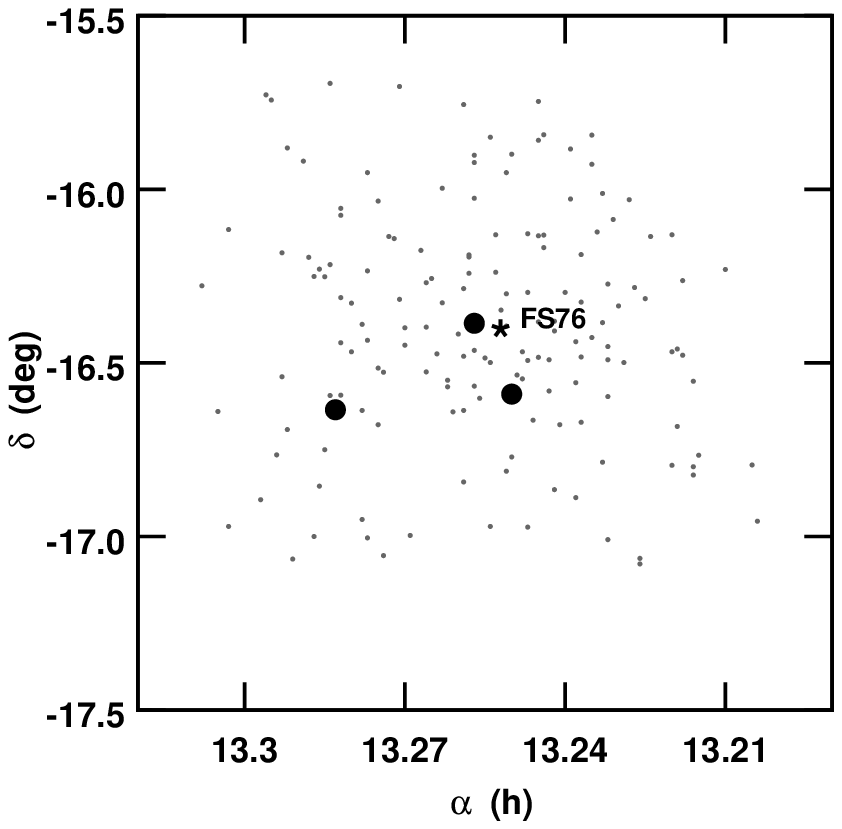"}
\caption{Position of FS76 (asterisk) inside the NGC5044 group. Small
dots indicate the positions of the 160 group members listed in
NED. Large dots indicate the positions of galaxies brighter than $M_B
= -20$. \label{catFS76ps}}
\end{figure}

Before discussing our observations, we first focus on the environments
of FS76 and FS373. Using ROSAT observations of the X-ray emitting gas
in the NGC5044 group, which is dominated by a single central giant
elliptical, NGC5044, \cite{da94} derive a total gravitating mass of $M
\approx 1.6 \times 10^{13}h^{-1}_{50} M_\odot$ within a radius of
250$h^{-1}_{50}$~kpc, corresponding to $M/L_B \approx 130h_{50}$ in
solar units and a galaxy velocity dispersion $\sigma_{\rm gal}=
330$~km/s. The mean systemic velocity of the NGC5044 group is $v_{\rm
sys}=2549$~km/s, according to NED. The elongated NGC3258 group is
dominated by two giant ellipticals:~NGC3258 and NGC3268. ASCA
observations of the X-ray emitting gas in the NGC3258 group by
\cite{pe97} lead to a total mass estimate $M \approx 2.3 \times
10^{13}h^{-1}_{50} M_\odot$ within a radius of 240$h^{-1}_{50}$~kpc,
corresponding to $M/L_B \approx 240h_{50}$ and a galaxy velocity
dispersion $\sigma_{\rm gal} \approx 400$~km/s. The mean recession
velocity of the group is $v_{\rm sys}=2848$~km/s, according to
NED. The position of FS373 in the outskirts of the NGC3258 group is
indicated in Fig. \ref{catFS373ps}. FS76 on the other hand has a
position very close to the center of the group (in projection), at a
projected distance of 50~kpc west of NGC5044 (Fig. \ref{catFS76ps}).

\subsection{Kinematics}
\begin{figure*}
\vspace*{9.7cm}
\special{hscale=80 vscale=80 hsize=700 vsize=400
hoffset=25 voffset=-55 angle=0 psfile="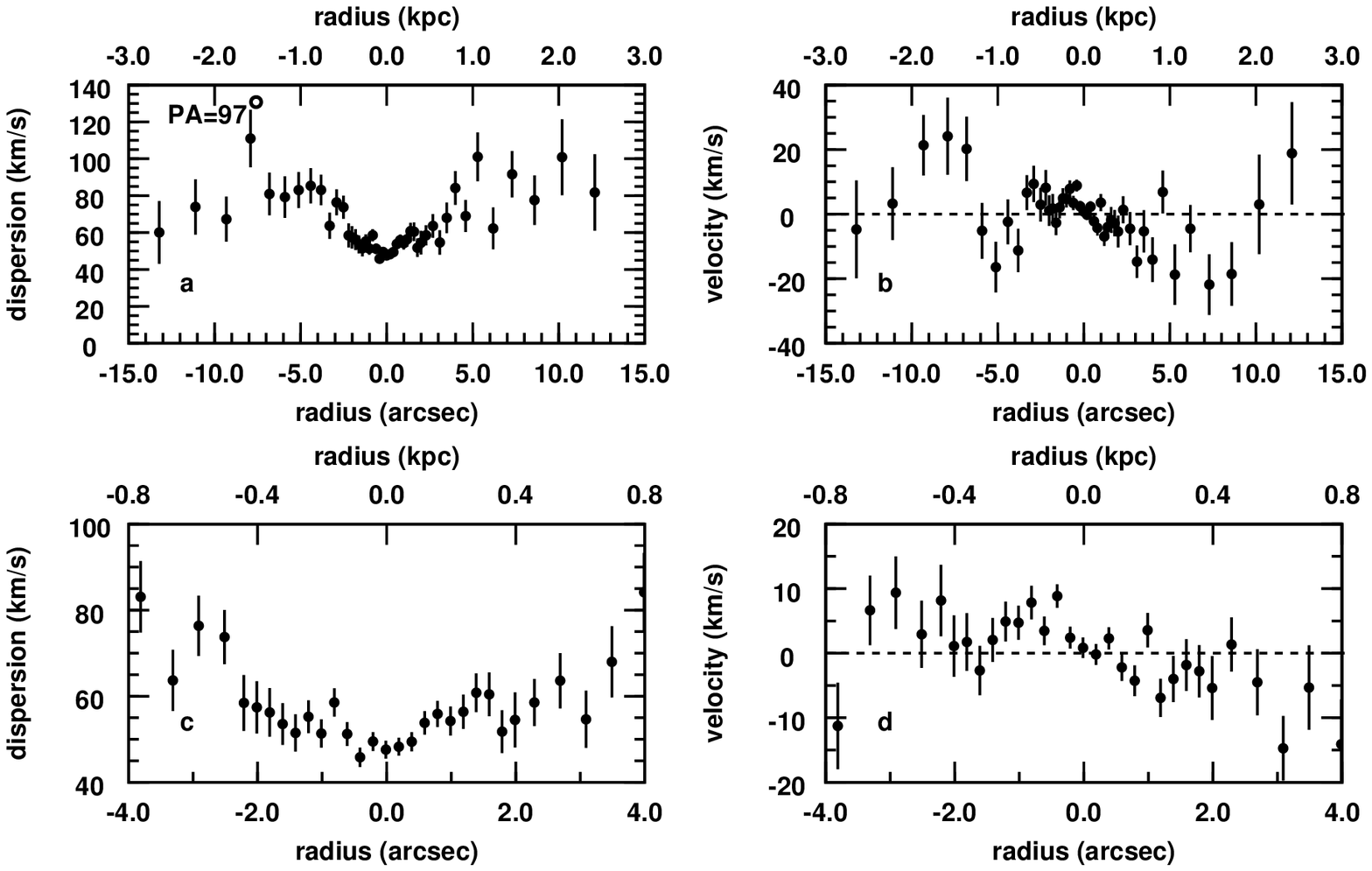"}
\caption{Major-axis kinematics of FS373. (a)~velocity dispersion,
(b)~mean velocity. The bottom panels zoom in on the central region
around the KDC to show the velocity dispersion (c) and mean velocity
(d) in more detail. The linear distance scale in kiloparsecs is
indicated above the top panels (assuming $H_0=70$~km/s/Mpc). The slit
position on the sky is indicated in panel (a). FS373 shows a
pronounced bump in the mean velocity profile around $\sim 1''$ and
falls to zero at $\sim 2''$ before rising again; this is
characteristic of a KDC. The dispersion has a central depression,
within the inner arcsecond, of $\sim 7$~km/s, which could hint at the
presence of a cold stellar disk. \label{kin1ps}}
%\end{figure*}
%\begin{figure*}
\vspace*{9.7cm}
\special{hscale=80 vscale=80 hsize=700 vsize=400
hoffset=25 voffset=-55 angle=0 psfile="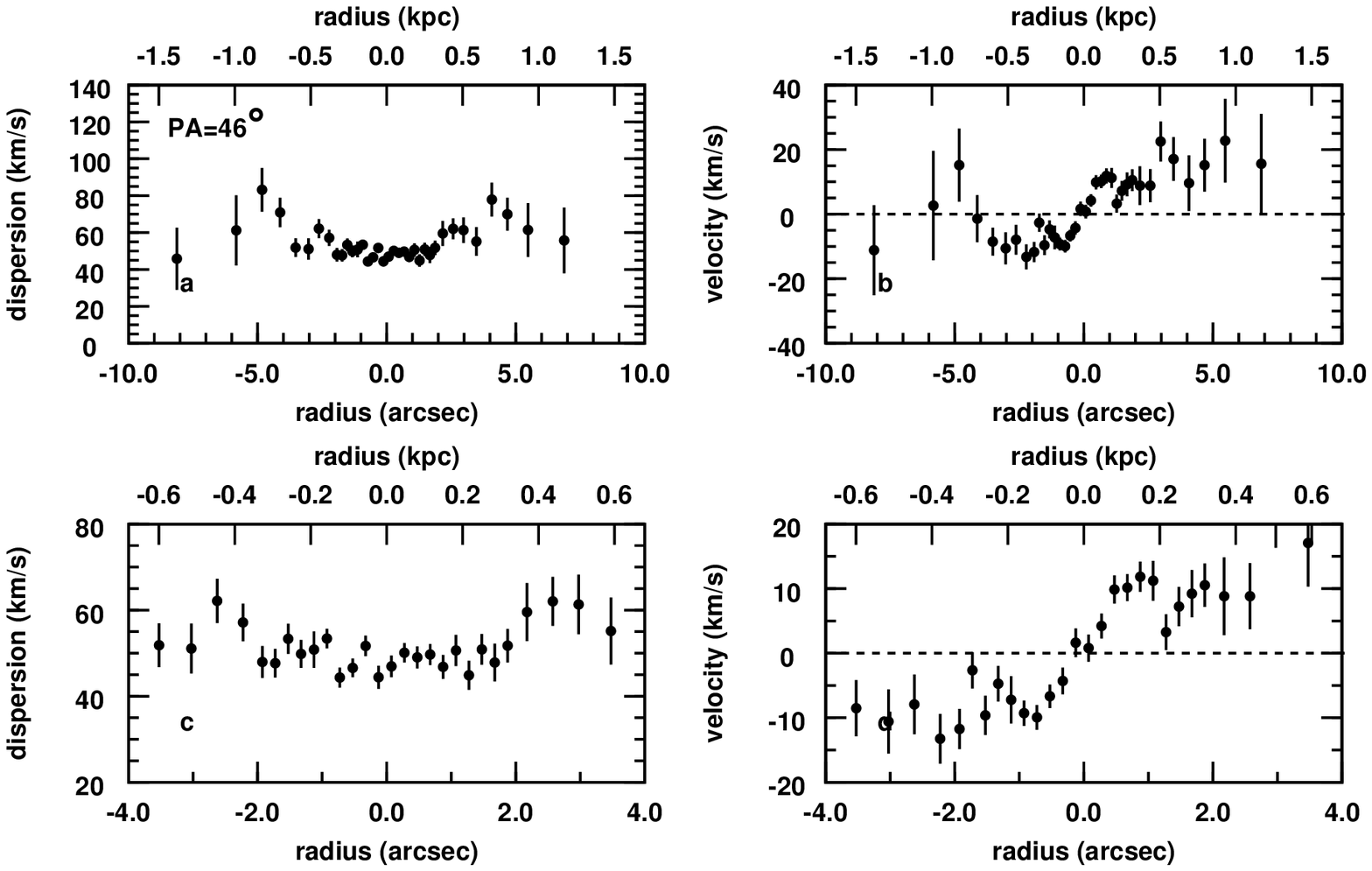"}
\caption{Major-axis kinematics of FS76. Same layout as in
Fig. \ref{kin1ps}. The giant elliptical NGC5044 is at a projected
distance of 50~kpc towards the east (i.e. positive radii are closest
to NGC5044). FS76 shows a pronounced bump in the mean velocity profile
around $\sim 0.8''$. The velocity dispersion remains approximately
flat at $\sim 50$~km/s within the inner $2''$ and rises rather sharply
outside this radius, indicative of a cold stellar
disk. \label{kin2ps}}
\end{figure*}
\begin{figure*}
\vspace*{17.5cm}
\special{hscale=70 vscale=70 hsize=800 vsize=800
hoffset=-15 voffset=-45 angle=0 psfile="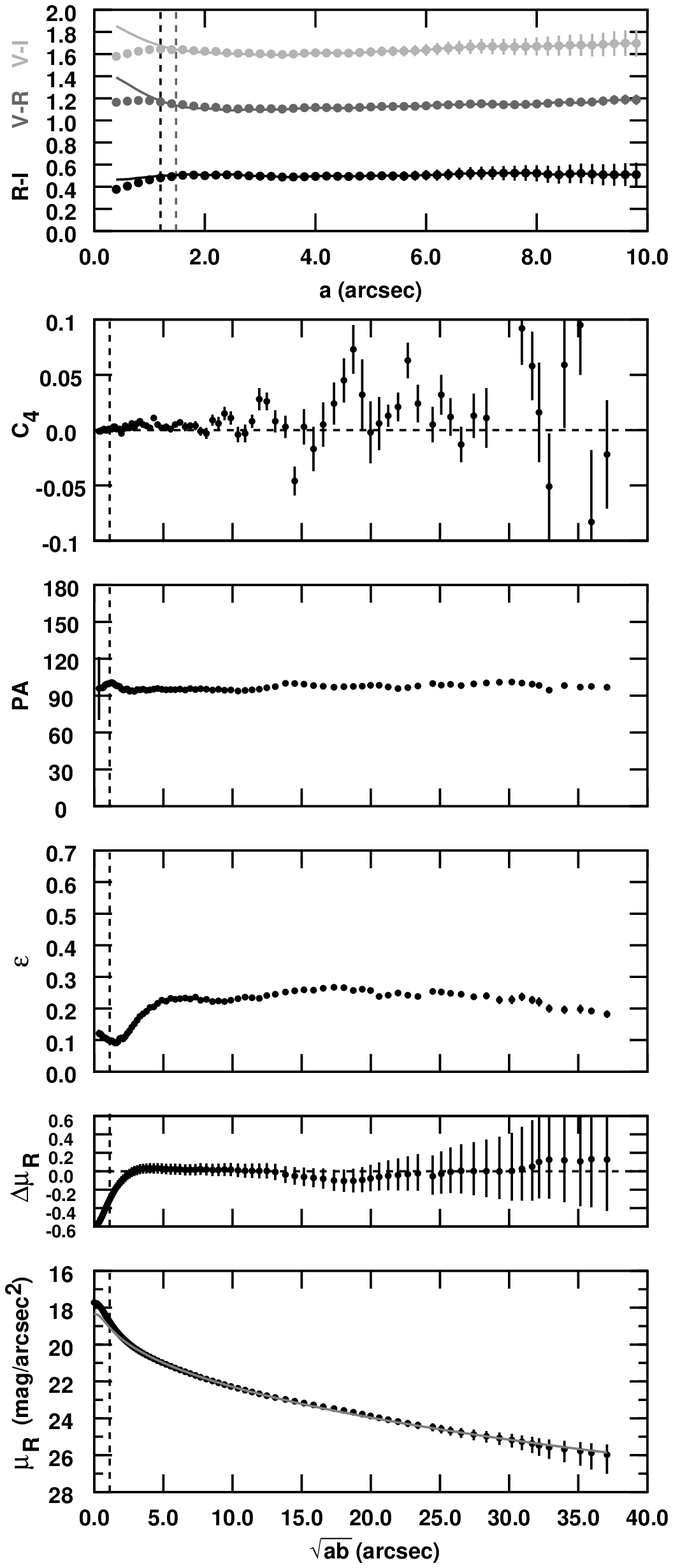"}
\special{hscale=70 vscale=70 hsize=800 vsize=800
hoffset=220 voffset=-45 angle=0 psfile="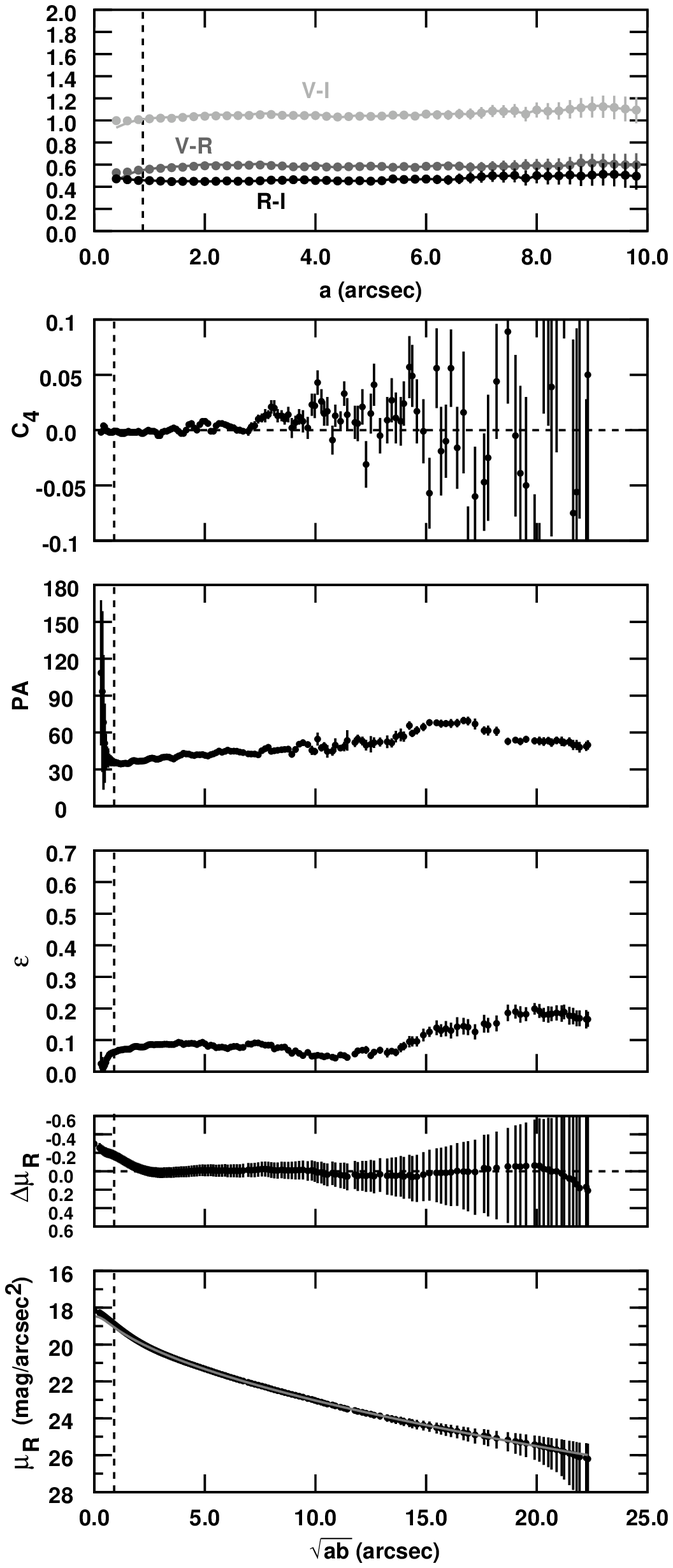"}
\caption{Photometry of FS373 (left column), photometry of FS76 (right
column). From the bottom up:~R-band surface brightness, $\mu_R$,
overplotted with the best fitting seeing-convolved S\'ersic profile
$\mu_R^S$ in grey; the residual from the S\'ersic fit, $\Delta \mu_R
=\mu_R - \mu_R^S$;  ellipticity, $\epsilon = 1-b/a$; position
angle, PA; and $C_4$, measuring the deviations of the isophotes from a
pure elliptical shape (these photometric quantities have all been
derived from the R-band image). Vertical dotted lines indicate the
seeing FWHM. The central nucleus, which sits in the center of the
outer isophotes, has been used as zero-point for the distance scale
(we use the geometric mean of the semi-major and semi-minor axes of
each isophotal ellipse, $\sqrt{ab}$, as radius). The positivity of the
$C_4$ parameter indicates that FS373 has disky isophotes. Top
panels:~R$-$I, V$-$R, and V$-$I color profiles as a function of
major-axis distance $a$. For each color index, the image with the best
seeing was convolved with a Gaussian to match the seeing of the other
image (vertical dotted lines indicate where the seeing correction
becomes important, with the V$-$R and V$-$I colors having a worse
seeing (grey line) than the R$-$I color (black line) in the case of
FS373). In the case of FS373, the V-band image suffers from variable
extinction by thin cirrus, affecting the V-band surface brightness
level but not the shape of the surface-brightness profile nor that of
the color profiles. The uncorrected color profiles are drawn in full
lines. The seeing correction is quite substantial, especially in the
profiles involving the V-band image (0.2-0.3~mag). For FS76, the
seeing corrections are small; 0.05 at most.
\label{phot373}}
\end{figure*}
The systemic velocity of FS373, $v_{\rm sys}=2415 \pm 1$~km/s, lies
within 1~$\sigma_{\rm gal}$ of the mean recession velocity of the
NGC3258 group, confirming it as a group member. Judging from panels
(b) and (d) in Fig. \ref{kin1ps}, FS373 has a complex mean velocity
profile. The KDC's solid body rotation dominates the central
kinematics and reaches a peak velocity of $6 \pm 2$~km/s around 200~pc
(or 1/8~$R_{\rm e}$) while outside the KDC the velocity rises to $20
\pm 5$~km/s at about 1~$R_{\rm e}$. The velocity changes sign again
around 2.2~kpc$\,\, \approx 1.5\,R_{\rm e}$. The velocity dispersion,
presented in panels (a) and (c) in Fig. \ref{kin1ps}, rises from $48
\pm 2$~km/s up to $90 \pm 12$~km/s at 1~$R_{\rm e}$ and remains flat
beyond that radius (the outermost data points suggest that the
dispersion may decline further out). Within the inner arcsecond, the
velocity dispersion shows a pronounced drop of about 7~km/s,
suggestive of the presence of a dynamically cold sub-component.

FS76 has a systemic velocity $v_{\rm sys}=2726 \pm 2$~km/s that places
it within $1~\sigma_{\rm gal}$ of the mean NGC5044 group
velocity. NGC5044 has almost the same systemic velocity:~$v_{\rm
sys}^{\rm N5044}=2704 \pm 33$~km/s (\cite{vsys5044}). The kinematics of FS76,
presented in Fig. \ref{kin2ps}, are equally intriguing as those of
FS373. The major axis velocity rises rapidly to $10\pm 3$~km/s at a
radius of about 200~pc$\, \, \approx 1/4 \,R_{\rm e}$ while the bulk
rotation flattens off at $15\pm 6$~km/s beyond $ 600$~pc$\,\, \approx
1 \, R_{\rm e}$. The velocity dispersion is virtually constant at $50
\pm 3$ km/s within the inner $2'' \approx 0.5\,R_{\rm e}$ and rises up
to $75 \pm 8$~km/s at $1\,R_{\rm e}$. Surprisingly, the velocity
dispersion declines beyond that radius and drops to $50 \pm 15$~km/s
at $2\,R_{\rm e}$. There is a hint of minor-axis rotation ($<5$~km/s),
which could be attributed to an inclined embedded disk
(\cite{dr1}). Outside 1$\,R_{\rm e}$, the velocity and velocity
dispersion profiles are asymmetric, suggestive of a past gravitational
interaction, most likely with NGC5044.

\subsection{Photometry and CaT$^*$ index}
\begin{table*}
\caption{Basic photometric parameters of FS76 and
FS373:~reddening-corrected V magnitude $m_V^0$, exposure time $t_V$
(seconds), seeing FWHM$_V$ (arcsec), and their R and I band analogs,
and the R-band effective radius $R_{\rm e}$. \label{tab1}}
\begin{center}
\begin{tabular}{|c|ccc|ccc|ccc|c|} \hline
	& $m_V^0$ & $t_V$ & FWHM$_V$ & $m_R^0$ & $t_R$ & FWHM$_R$ &
$m_I^0$ & $t_I$ & FWHM$_I$ & $R_{\rm e}$~($''$/kpc) \\ \cline{1-11}
FS373 & / & 720 & 1.5$''$ & 14.17 & 360 & 1.2$''$ & 13.76 & 270 &
1.1$''$ & 7.9/1.57 \\ FS76 & 15.44 & 180 & 0.6$''$ & 14.84 & 220 &
0.9$''$ & 14.35 & 300 & 0.8$''$ & 4.4/0.77 \\\cline{1-11}
%FS76 & 15.44 & 14.84 & 14.35 &
%4.4/0.77 & 1.85 \\ \cline{1-6}
\end{tabular}
\end{center}
\end{table*}
\begin{table}
\caption{R-band S\'ersic parameters of FS373 and FS76:~extrapolated
central surface brightness $\mu_{\rm R,0}$ (at zero seeing),
scale-length $r_0$~(arcsec), and shape-parameter $n$. \label{tab2}}
\begin{center}
\begin{tabular}{|c|ccc|} \hline
	& $\mu_{\rm R,0}$ & $r_0$ & $n$ \\ \cline{1-4} FS373 & $16.55
\pm 0.16$ & $0.11 \pm 0.02$ & $2.71 \pm 0.08$ \\ FS76 & $17.58 \pm 0.03$
& $0.50 \pm 0.02$ & $1.85 \pm 0.02$ \\ \cline{1-4}
\end{tabular}
\end{center}
\end{table}
We measured the surface-brightness profile, position angle, and
ellipticity $\epsilon=1-b/a$ of FS373 and FS76 as a function of the
geometric mean of major and minor axis distance, denoted by $a$ and
$b$ respectively. These were obtained using our own
software. Basically, the code fits an ellipse through a set of
positions where a given surface brightness level is reached. The shape
of an isophote, relative to the best fitting ellipse with semi-major
axis $a$ and ellipticity $\epsilon$, is quantified by expanding the
intensity variation along this ellipse in a fourth order Fourier
series with coefficients $S_4$, $S_3$, $C_4$ and $C_3$~:
\begin{eqnarray}
I(a,\theta) &=& I_0(a) \left[ 1 + C_3(a) \cos(3\theta)+ C_4(a)
\cos(4\theta)) \right.+ \nonumber \\ && \left. S_3(a)\sin(3\theta))+
S_4(a) \sin(4\theta) \right]. \label{ith}
\end{eqnarray}
Here, $I_0(a)$ is the average intensity of the isophote and the angle
$\theta$ is measured from the major axis. The basic photometric
parameters of the two galaxies are presented in Table \ref{tab1}. The
surface-brightness, position angle, ellipticity, and $C_4$ profiles
are presented in Fig. \ref{phot373}.

We fitted a seeing-convolved S\'ersic profile to the R-band surface
brightness profile $\mu_{\rm R}(r) = \mu_{\rm R,0} +
1.086(r/r_0)^{1/n}$, or equivalently $\mu_{\rm R}(r) = \mu_{\rm R,0} +
2.5 b_n (r/R_e)^{1/n}$, with $r_0$ the scale-length, $R_e$ the
half-light radius, $b_n \approx 0.848 n - 0.142$, radius
$r=\sqrt{ab}$, measured from the central nucleus or intensity
peak. The seeing characteristics were estimated from about 10 stars in
each image. We simply minimized the quadratic difference between the
observed and the seeing-convolved S\'ersic profile using a non-linear
minimization routine. The formal error bars are approximated by the
diagonal elements of the estimated covariance matrix of this
non-linear problem. The results of this fit are presented in
Fig. \ref{phot373} and Table \ref{tab2}. Since both galaxies have a
central brightness peak that cannot be fitted with a S\'ersic profile,
the inner 2$''$ were excluded from the fit. The value $n=4$
corresponds to a de Vaucouleurs profile, typical for massive
ellipticals; $n=1$ corresponds to the diffuse exponential profile,
typical for the faintest dwarf ellipticals. The $n$ values found by us
agree with the dE classification of these galaxies.

\begin{figure}
\vspace*{5cm}
\special{hscale=55 vscale=55 hsize=700 vsize=230
hoffset=-27 voffset=-195 angle=0 psfile="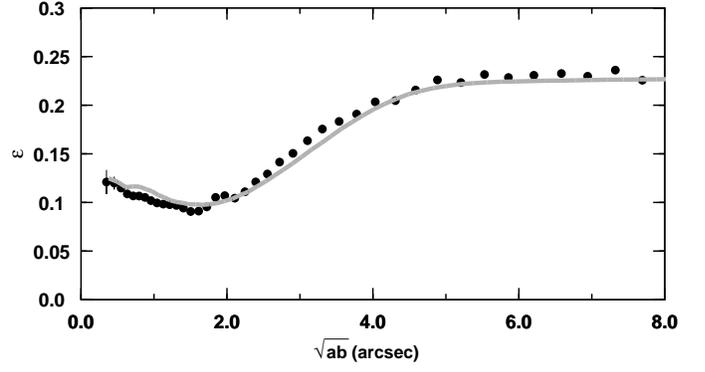"}
\caption{Simulated ellipticity profile of FS373. A flat disk with
$\epsilon=1-b/a=0.9$ was added to a host galaxy with a flattening
varying smoothly between $\epsilon=0$ in the very center and
$\epsilon=0.23$ at 6$''$ and beyond. This model was then convolved
with a Gaussian to simulate FWHM=1.2$''$ seeing conditions and sampled
with $0.2''\times0.2''$ pixels in order to reproduce the FORS2 CCD
sampling. Both the position angle and the surface brightness profile
match the R-band characteristics of FS373. Without attempting to
reconstruct the appearance of FS373 in detail, this toy model (grey
full line) is able to reproduce the central rise of the observed
ellipticity profile (black dots). \label{disk373}}
\end{figure}

FS373 has a systematically positive $C_4$ and its isophotes are
consequently slightly disky. Moreover, the ellipticity does not
decline all the way to the center but instead rises inwardly inside
the inner 2$''$. If the true flattening were constant all the way to
the center, convolution with a circularly symmetric Gaussian seeing
profile would make the inner isophotes steadily rounder as one
approaches the center. In order to understand the observed behavior of
the ellipticity profile, we calculated the appearance of a galaxy with
a flattening varying smoothly between $\epsilon=0$ in the very center
and $\epsilon=2.3$ at 6$''$ and beyond, to which an exponential disk
with $\epsilon=9$ and radial scale-length $h_R=0.3''$ is added. This
model was then convolved with a Gaussian to simulate 1.2$''$~FWHM
seeing conditions and sampled with $0.2''\times0.2''$ pixels in order
to reproduce the FORS2 CCD sampling. Both the position angle and the
surface brightness profile match the R-band characteristics of
FS373. Without attempting to reconstruct the appearance of FS373 in
detail, this toy model (grey full line in Fig. \ref{disk373}) is able
to reproduce the central rise of the observed ellipticity profile.

\begin{figure}
\vspace*{6.7cm}
\special{hscale=70 vscale=70 hsize=700 vsize=230
hoffset=0 voffset=-240 angle=0 psfile="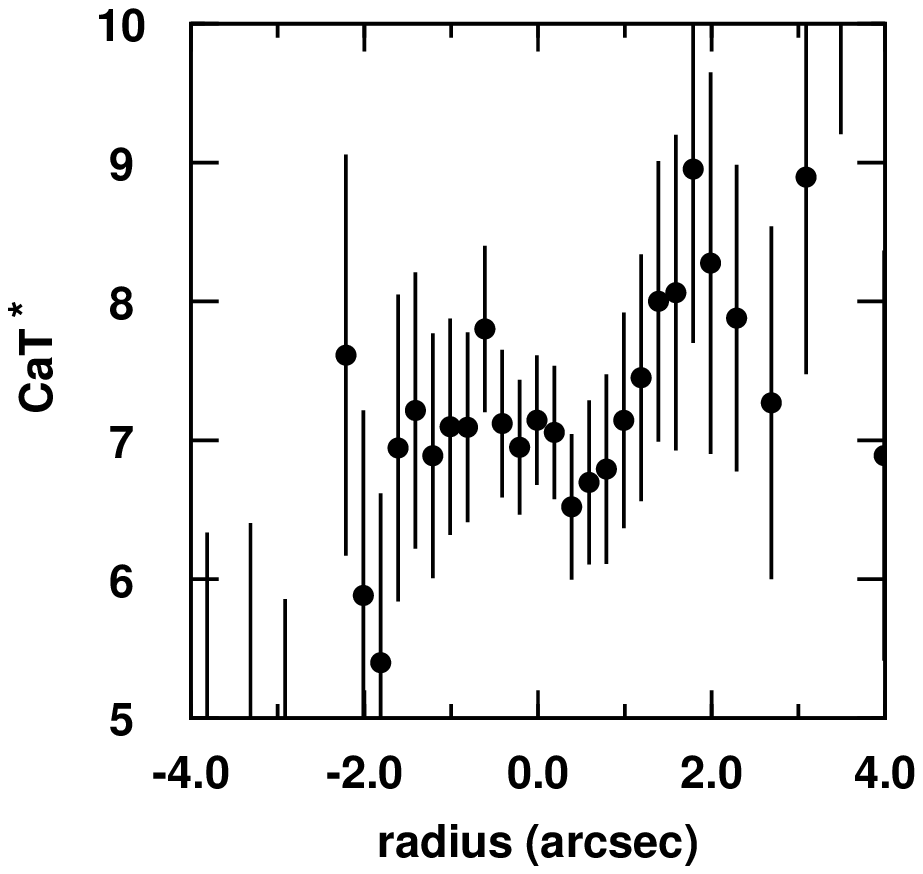"}
\caption{The strength of the Ca{\sc ii} triplet lines, corrected for
contamination by the Pa lines, as measured by the CaT$^*$ index along
the major axis of FS373. \label{ca373}}
%\end{figure}
%\begin{figure}
\vspace*{6.7cm}
\special{hscale=70 vscale=70 hsize=700 vsize=230
hoffset=0 voffset=-240 angle=0 psfile="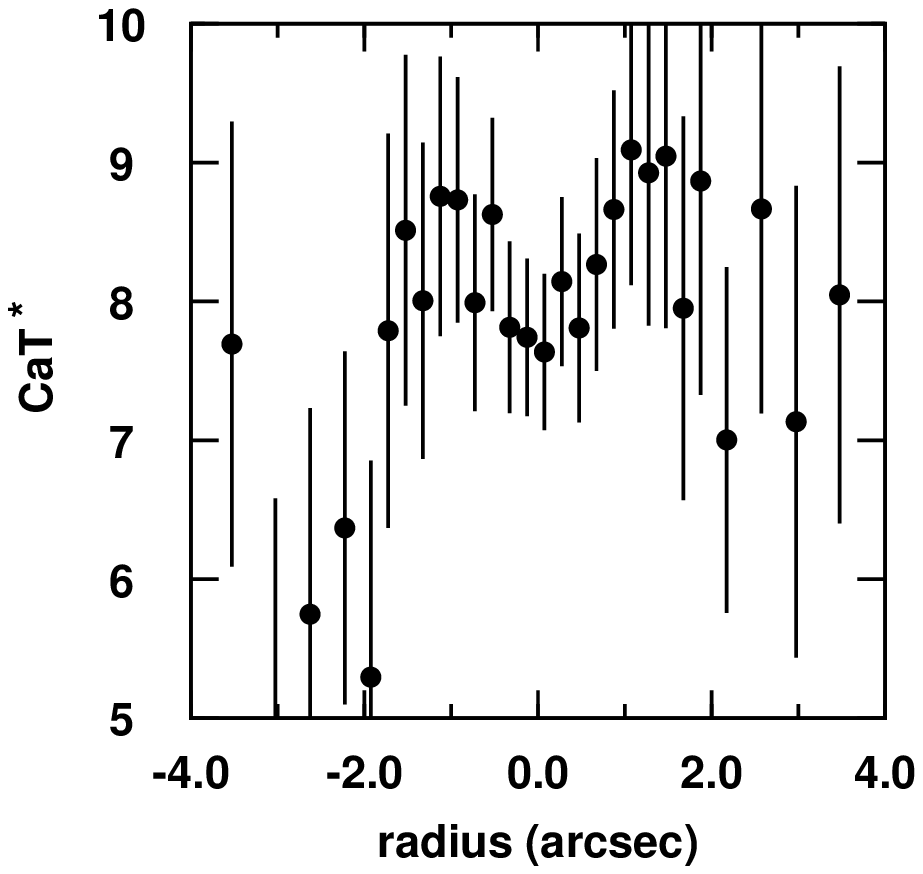"}
\caption{The strength of the Ca{\sc ii} triplet lines, corrected for
contamination by the Pa lines, as measured by the CaT$^*$ index along
the major axis of FS76. \label{ca76}}
\end{figure}

The R$-$I, V$-$R, and V$-$I color profiles of FS373 are presented in
Fig. \ref{phot373} (top panel). These are constructed by first fitting
a cubic spline to the various surface brightness profiles as a
function of radius $r = \sqrt{ab}$ so they can all be evaluated at any
given position. The errorbars take into account photon shot noise and
sky-subtraction uncertainties. For each color, the image with the best
seeing was convolved with a two-dimensional Gaussian to match the
seeing of the other image (see Table \ref{tab1} for the basic
photometric data). For FS373 this correction towards equal seeing was
quite substantial ($0.10-0.25$~mag near the center). The V-band image
suffers from variable extinction by thin cirrus, affecting the surface
brightness level but not the {\em shape} of the surface-brightness
profile nor that of the color profiles. Outside the central region,
FS373 has slightly rising color profiles, an effect best seen in the
V$-$I profile. This outward reddening has been observed in many dEs
and is usually interpreted as a metallicity effect (\cite{my}). Inside
the inner 2$''$, all color profiles show a pronounced bump before
declining inward, although one has to take into account the
uncertainty that comes with the large seeing correction that was
applied. Nonetheless, this behavior correlates with the strength of
the Ca{\sc ii} lines as measured by the CaT$^*$ index
(\cite{ce01}). CaT$^*$ first rises outwards but appears to decline
outside a radius of 2$''$ until the index can no longer be measured
reliably (amongst other due to residuals of the ubiquitous strong sky
emission lines that make it difficult to properly measure the
continuum level). If CaT$^*$ traces metallicity in a composite stellar
population, which is still the subject of debate, then one would
indeed expect the KDC to be slightly redder than its immediate
surroundings.

Photometric parameters for FS76 are presented in Fig. \ref{phot373}
and Table \ref{tab1}. The isophotes of FS76 do not deviate
significantly from ellipses ($C_4 \approx 0$). Within the inner
$1.2''$, the position angle shows a pronounced twist. Color profiles
for this galaxy are shown in Fig. \ref{phot373}. The V$-$I color
varies almost linearly with radius, going from V$-$I=1.0~mag at
$0.5''$ to V$-$I=1.1~mag at $10''$. This outward reddening is also
visible in the R$-$I profile, although less obvious, going from
R$-$I=0.45~mag at $2.0''$ to R$-$I=0.50~mag at $10''$. The inner $2''$
appear to be slightly redder again in R$-$I but bluer in V$-$R and
V$-$I. The CaT$^*$ index profile, presented in Fig. \ref{ca76}, shows
approximately the same behavior as that of FS373:~the strength of the
Ca{\sc ii} lines rises outwardly out to about 1.5$''$ and declines
further out. The higher CaT$^*$ value near the position where the KDC
dominates the kinematics may either point to a slightly more metalrich
(a few tens of a dex) or an older stellar population. Taken at face
value, the colors and line-strengths are roughly consistent with a
metalpoor ([Fe/H]$\,\,\approx -0.7$) and old ($T>6$~Gyr) star
population.  (\cite{va,va03}).

Clearly, imaging at much higher spatial resolution, e.g. with HST, is
required to obtain more reliable colors close to the center and to
corroborate the presence of a disky subcomponent. Also, spectroscopy
in a wavelength region better suited for constraining the mean age and
metallicity of the stellar population is prerequisite to fully
understand the origin of these KDCs. Finally, it should be noted that
neither galaxy shows any dust features.

\section{Dynamical models} \label{dyn}

The internal dynamics of a steady-state axisymmetric stellar system
are described by a gravitational potential $\psi(\varpi,z)$, that
determines the stellar orbits, and the distribution function (DF)
$F(\vec{r},\vec{v})\,d\vec{r}\,d\vec{v}$, which gives the number
density of stars in phase space ($(\varpi,z)$ are cylindrical
coordinates). Roughly speaking, the DF distributes the stars over all
possible orbits. According to the Jeans theorem, the DF can be written
as a function of the isolating integrals of motion. An axisymmetric
potential generally allows only two such integrals; the binding energy
$E$ and the $z$-component of the angular momentum $L_z$. More freedom
to distribute stars over phase space can be gained if a third integral
of motion exists. Therefore, we work in spheroidal coordinates and
approximate the gravitational potential by a St\"ackel potential which
allows the existence of a third integral, denoted by $I_3$. Round
galaxies are modeled more efficiently in a spherical geometry, in
which case the DF is conveniently taken to be a function of $E$,
$L_z$, and $L$, the total angular momentum. The $L_z$-dependence of
the DF allows the construction of rotating and slightly flattened
stellar systems. FS373, which has a significantly flattened appearance
on the sky, is modeled using an axisymmetric St\"ackel potential while
FS76 is treated as a system with a spherical gravitational
potential. Of course, a galaxy with a KDC need not be axisymmetric but
in the context of equilibrium models such an assumption is inevitable;
also, the isophotes do not indicate that these galaxies are
significantly non-axisymmetric.

A detailed account of the method we employed to construct the
spheroidal coordinate system and St\"ackel potential that give the
best fit to a given axisymmetric potential can be found in
\cite{dz88,de96}, and \cite{deb01}. In brief, we deproject the
observed surface brightness distribution, derived from an I-band
image, assuming the galaxy to be axisymmetric and viewed edge-on. The
total mass density, including dark matter, is parameterized as the
spatial luminosity density multiplied by a spatially varying
mass-to-light ratio
\begin{equation}
\Upsilon(\varpi,z) = A \left(1 + B \sqrt{\varpi^2 + (z/q)^2}\right),
\end{equation}
with the parameters $A$ and $B$ to be estimated from the data and $q$
the axis ratio of the luminosity density distribution. The
gravitational potential is obtained by decomposing the total mass
density in spherical harmonics. Finally, we fit a St\"ackel potential
to this gravitational potential. For a spherical galaxy ($q=1$), the
deprojection of the surface brightness profile reduces to solving an
Abel's integral equation (\cite{bt}, eq. (1B-57b)). The gravitational
potential follows from \cite{bt}, eq. (2-22).

For a given potential, we wish to find the DF that best reproduces the
kinematical information. The DF is written as a weighted sum of basis
functions $F = \sum_i c_i \, F_i$ and the coefficients $c_i$ are
determined by minimizing the quantity
\begin{equation}
\chi^2 = \sum_l \left( \frac{ {\rm obs}_l - \sum_i c_i \,{\rm
obs}_{l,i} }{\sigma_l} \right)^2,
\end{equation}
with ${\rm obs}_l$ an observed data point, $\sigma_l$ the 1-$\sigma$
errorbar on that data point, and ${\rm obs}_{l,i}$ the corresponding
value calculated from the $i^{\rm th}$ basis function, subject to the
constraint that the DF be positive everywhere in phase space. For a
spherical potential, we used the basis functions $F_i(E,L) =
E^{\alpha_i} L^{\beta_i}$ and $F_j(E,L_z) = E^{\alpha_j}
L_z^{\beta_j}$, with integer powers, to construct the DF
(\cite{deb04}). The three-integral DF in a St\"ackel potential on the
other hand was built with basis functions of the form $F_i(E,L_z,I_3)
= E^{\alpha_i} L_z^{\beta_i} I_3^{\gamma_i}$, with integer powers
(\cite{de96}). In the case of FS76, the models are fitted directly to
the observed major and minor axis spectra (\cite{dr0}), making the
best possible use of all the kinematical information contained in the
spectra. The axisymmetric models for FS373 are much more complex and
computationally time-consuming and we opted to use the observed
surface brightness distribution and the velocity dispersion and mean
velocity profiles along both major and minor axis (i.e. the central
second and first order moments of the LOSVDs, {\em not} the Gaussian
parameters) as kinematical input for the modeling code. As a further
constraint, we used the central fourth order moment of the LOSVDs,
calculated from the kinematic parameters up to $h_4$. We use a
Quadratic Programming technique to find the optimal values for the
$c_i$ using a library of hundreds of linearly independent basis
functions, making this method essentially non-parametric. This is
repeated for about 100 different ($A,B$)-pairs, covering the relevant
part of parameter space. This allows us to define the range of models
(and hence mass distributions) that are consistent with the data and
to determine which model gives the best fit to the data.

\begin{figure}
\vspace*{9.75cm}
\special{hscale=47 vscale=47 hsize=700 vsize=330
hoffset=-14 voffset=-25 angle=0 psfile="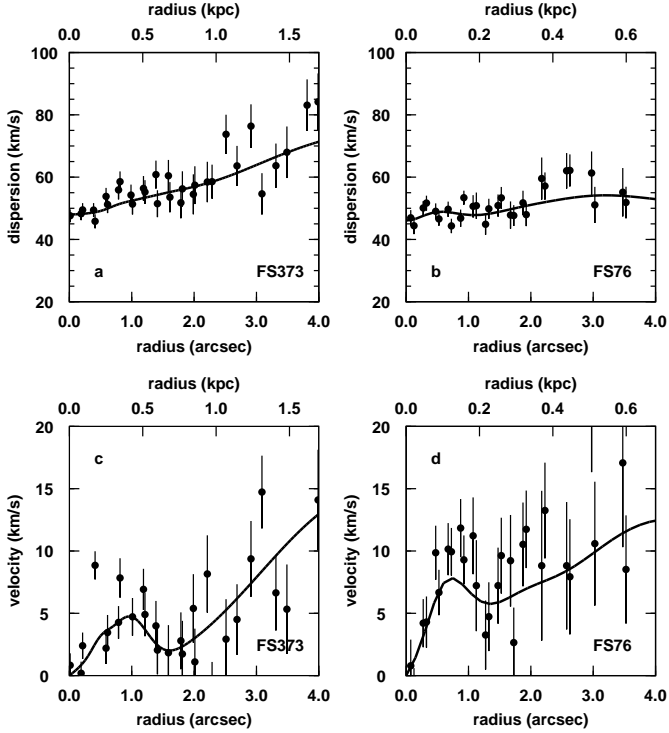"}
\caption{Major axis velocity dispersion and mean velocity of FS373
(panels (a) and (c), respectively) and of FS76 (panels (b) and (d)),
folded around the center of the galaxy. The corresponding profiles of
the best fit models are plotted with a full line. The models for FS373
were fitted to these kinematics. On the other hand, the models for
FS76 were fitted directly to the spectra and are independent from the
kinematics determined by a Gauss-Hermite fit to the LOSVDs. Still, the
model agrees excellently with the kinematics, as it should of
course. \label{modelps}}
\end{figure}

\begin{figure}
\vspace*{4.7cm}
\special{hscale=52 vscale=52 hsize=700 vsize=360
hoffset=-44 voffset=-50 angle=0 psfile="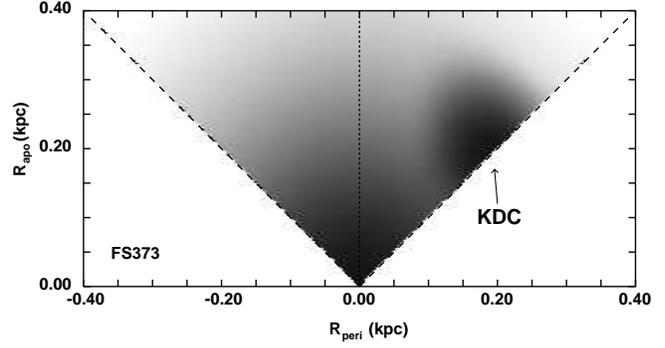"}
\caption{The distribution function of FS373 in the equatorial plane
 in turning-point space. The pericenter distance, $R_{\rm
peri}$, has the same sign as the $z$-component of the angular
momentum, $L_z$. Circular orbits have $R_{\rm apo} = \pm R_{\rm peri}$
(dashed lines); radial orbits are characterized by $R_{\rm peri}=0$
(dotted line). The KDC is visible in phase-space as an excess of stars
on near-circular orbits, indicated by an arrow. \label{mod373}}
\end{figure}

\begin{figure}

\vspace*{4.7cm}
\special{hscale=52 vscale=52 hsize=700 vsize=360
hoffset=-44 voffset=-50 angle=0 psfile="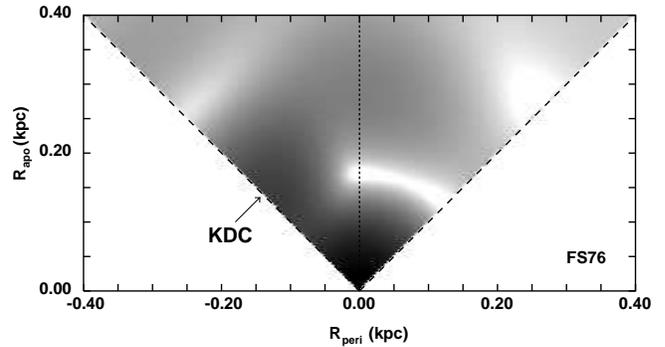"}
\caption{The distribution function of FS76 in the equatorial plane in
turning-point space. The KDC is visible in phase-space as an excess of
stars on near-circular orbits. Its locus is indicated by an
arrow. \label{mod76}}
\end{figure}

The major-axis velocity dispersion and mean velocity profiles of the
best models for FS373 and FS76 are compared with the observed
kinematics in Fig. \ref{modelps}. From these best models, we estimate
the mass within a 1.5~$R_{\rm e}$ sphere of FS76 at
$2.9^{{+0.5}}_{{-1.7}} \times 10^9 M_\odot$
%$2.9^{^{+0.5}}_{^{-1.7}} M_\odot$
(at the 90\% confidence level) and that of FS373 at
$6.2^{{+2.3}}_{{-1.3}} \times 10^9 M_\odot$. Both galaxies require the
presence of a dark matter halo, albeit not a very massive one, in
order to reproduce the observed outwardly rising velocity dispersion
profiles. For FS76 we find a B-band mass-to-light ratio $M/L_B =
7.8^{{+1.3}}_{{-4.6}} M_\odot/L_{B,\odot}$ while for FS373 we find
$M/L_B = 7.4^{{+2.8}}_{{-1.5}} M_\odot/L_{B,\odot}$. These models
agree very well with the data and reproduce the central dip in the
velocity dispersion, the bump in the velocity profile, and, in the
case of FS373, the inward rise of the ellipticity profile. In both
galaxies, this requires the presence of a fast rotating cold
subcomponent in the models. The DFs of the best models for FS373 and
FS76 are presented in Figs. \ref{mod373} and \ref{mod76},
respectively. We plot the DF in the equatorial plane in turning-point
space. Each orbit in this plane is labeled uniquely by its pericenter
distance $R_{\rm peri}$ and apocenter distance $R_{\rm apo}$ if
$R_{\rm peri}$ is given the same sign as $L_z$. Circular orbits lie on
two straight lines with $R_{\rm apo} = \pm R_{\rm peri}$. Radial
orbits lie on the vertical line with $R_{\rm peri}=0$. In both galaxy
models, an excess phase-space density of stars on near-circular
orbits, forming the KDC, is clearly visible. Moreover, the KDC is
obviously disjunct from the central nucleus or density cusp. Since the
KDCs form a distinct subcomponent within their host galaxies, the
stars that make up a KDC can be singled out of the DF and be studied
separately (especially in the case of FS373, it was very clear which
basis functions in the expansion of the DF formed the KDC). In order
to roughly estimate the stellar mass of the KDC, we assumed a stellar
mass-to-light ratio of $M/L_B = 2-4 M_\odot/L_{B,\odot}$, which agrees
with the observed colors and line-strengths. Thus, we find $M_{\rm
KDC} \approx 1-5 \times 10^7 M_\odot$ for both galaxies or a few
percent at most of the total mass.  The adopted $M/L_B$ is both
typical for a 10~Gyr old, metal-poor ($-1<$[Fe/H]$<-0.5$) stellar
population (which would agree with dEs being primordial stellar
systems) and for a 5~Gyr old, more metal-rich ($-0.5<$[Fe/H]$<0.0$)
stellar population (which would agree with dEs being harassed
late-type spirals that experienced a starburst) (\cite{wo94}).


\section{Discussion} \label{dis}

The key question is whether KDCs in dwarf elliptical galaxies are
produced the same way as in massive ellipticals. We explore two
possible avenues to KDC formation in dEs. The first is the merger
hypothesis, as in giant ellipticals; the second is the harassment
scenario, which posits that gravitational interactions play an
important role in the evolution of dEs. The analytical arguments given
below are strictly speaking only valid for {\em fast} and {\em
distant} encounters. An encounter between to galaxies, with masses
$M_1$ and $M_2$, qualifies as {\em distant} if, at closest approach,
the change in the potential energy of the pair is much smaller than
the initial orbital kinetic energy. In a {\em fast} encounter, the
relative velocity of the galaxies is much larger than the internal
stellar velocities. This translates into the following constraints on
the impact parameter $b$ and the internal velocity dispersion
$\sigma_{\rm int}$~:
\begin{equation}
b > \frac{2G}{V_{\rm rel}^2}(M_1+M_2); \,\,
\sigma_{\rm gal} > \sigma_{\rm int},
\end{equation}
with $V_{\rm rel}$ the relative velocity of the interacting galaxies.
For $M_2 \approx 5 \times 10^9 M_\odot$, a typical dE mass, and $M_1 =
M_{\rm KDC} << M$, we find $b > 250 - 500$~pc for $V_{\rm rel} =
\sigma_{\rm gal} = 300 - 400$~km/s. Also, $\sigma_{\rm gal} >
\sigma_{\rm int}$. Hence, any non-penetrating encounter between a dE
and a much smaller dwarf galaxy classifies as a fast and distant
encounter (even if we take the dwarf galaxy to be originally 10 times
more massive than $M_{\rm KDC}$, the minimum impact parameter would
change by only 10\%). In the case of a giant elliptical with $M_2
\approx 5 \times 10^{11} M_\odot$ and $M_1 = M_{\rm dE} << M_2$, the
condition for a fast flyby becomes $b > 25 - 50$~kpc, again rather
unsensitive to $M_{\rm dE}$. In a group or cluster environment,
galaxies keep respectable distances of a few tens of kpc
(\cite{mkldo}). With this in mind, we can discuss possible mechanism
of producing KDCs in dEs.

\subsection{The merger hypothesis}

While the merger origin of KDCs in bright ellipticals is well
accepted, a number of facts argue against the merger hypothesis in the
case of dEs.

The change of the forward velocity of a galaxy with mass $M_1$ induced
by a fast, distant hyperbolic encounter with a galaxy with mass $M_2$
with a relative velocity $V_{\rm rel}$ is given by
\begin{equation}
\Delta V_{||} = - \frac{2 G^2}{b^2 V_{\rm rel}^3} M_2 (M_1+M_2),
\end{equation}
(\cite{sg}, \cite{bt}). The closer and the slower the encounter, the
more orbital energy is converted into internal (stellar) kinetic
energy. For an encounter between a typical $M_2 =5 \times 10^9
M_\odot$ dE and a $M_1 =5 \times 10^7 M_\odot$ dwarf galaxy with a
relative velocity $V_{\rm rel} = \sigma_{\rm gal} = 300$~km/s, $\Delta
V_{||}$ is very small (e.g. $\Delta V_{||} \sim 35$~km/s for a
collision with $b=1$~kpc). In the case of an encounter between a
$M_2=5 \times 10^{11} M_\odot$ elliptical and a $M_1=5 \times 10^9
M_\odot$ dE, on the other hand, the velocity change is
substantial~:~$\Delta V_{||} \sim V_{\rm rel}$, even for impact
parameters of a few tens of kiloparsecs. This suggests that a dE, in a
group or cluster environment, has virtually no chance of slowing down
and capturing another (smaller) dwarf galaxy, contrary to a more
massive elliptical galaxy. Hence, once the galaxy group or cluster is
in place, the chance of forming a KDC in a dE by a merger is
exceedingly small. Also, it is unclear how the merger scenario can
explain the complex velocity profile of FS373, particularly the
velocity changing sign around a radius of $12'' = 2.4$~kpc.

Alternatively, the merger could have taken place {\em before} the
group or cluster virialized, in an environment where relative
velocities were smaller than the present values. The low galaxy
density in such an environment argues against this idea. Also, it
remains to be seen, e.g. using high-resolution $N$-body simulations,
whether a KDC formed this way can survive the dE's falling into a
group or cluster and the subsequent gravitational interactions with
giant group or cluster members.

\subsection{The harassment scenario}

\begin{figure}
\vspace*{5.5cm}
\special{hscale=50 vscale=50 hsize=700 vsize=160
hoffset=-43 voffset=263 angle=-90 psfile="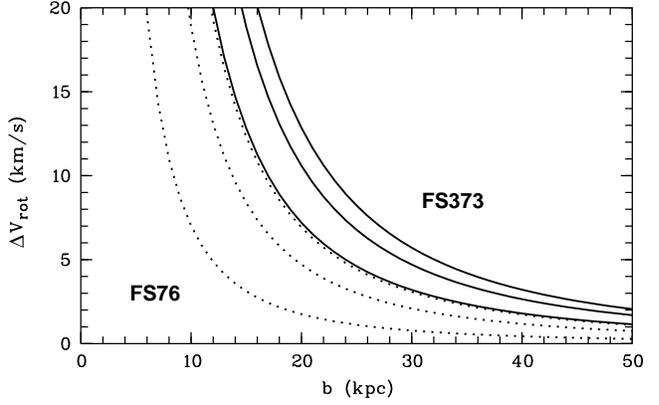"}
\caption{The estimated maximum change in the rotation velocity of a
dwarf galaxy after an encounter with a giant elliptical of mass $M_2 =
5 \times 10^{11} M_\odot$ as a function of impact parameter $b$ and
axis ratio $q$. Full lines:~FS373 (top to
bottom:~$q=0.3,\,0.5,\,0.7$), dotted lines:~FS76 (top to
bottom:~$q=0.5,\,0.7,\,0.9$). We used $V_{\rm rel}=300$~km/s in the
case of FS76 and $V_{\rm rel} =400$~km/s for FS373. Clearly,
interactions with impact parameters smaller than $b \sim 20$~kpc can
signifanctly change the streaming velocity of these slowly rotating
systems.
\label{vrot}}
\end{figure}
A plausible alternative is the spin-up of a dE's halo by fly-by
encounters with other galaxies. The impulse approximation and the
tensor virial theorem yield the following expression for the maximum
amount of angular momentum that can be transfered to a galaxy with
mass $M_1$ during an encounter with a galaxy with mass $M_2$:
\begin{equation}
\Delta J = 2 \frac{G M_2}{ b^2 V} I_{11} \left( 1-q_1^2 \right)
\end{equation}
with $q_1$ the axis ratio and $I_{11}$ a component of the inertial
tensor (\cite{ssk}). Using $\Delta J \sim M_1 R_{\rm e,1} \Delta
v_{\rm rot}$ to roughly estimate $\Delta v_{\rm rot}$, the maximum
possible change in the rotation velocity after re-virialization, this
translates into
\begin{equation}
\Delta v_{\rm rot} \sim \frac{2}{3} \frac{G M_2}{ b^2 V}
R_{\rm e,1} \left( 1-q_1^2 \right),
\end{equation}
with $R_{\rm e,1}$ the half-light radius of the galaxy with mass $M_1$
(we have used $I_{11} \sim M_1 R_{\rm e,1}^2/3$). This $\Delta v_{\rm
rot}$ is plotted in Fig. \ref{vrot} as a function of impact parameter
$b$ and axis ratio $q$. Clearly, interactions with impact parameters
small than $b \sim 20$~kpc can significantly alter the streaming
velocity of these slowly rotating systems, especially in the case of
FS373, which is apparently more flattened than FS76.

According to the harassment scenario, some dEs may stem from late-type
spirals, inflated by high-speed gravitational interactions
(\cite{mkldo}, \cite{ma}). During this metamorphosis, which in the
case of a late-type spiral orbiting a massive galaxy, e.g. in a small
group, takes only 2$-$3 pericenter passages, a disk and a spheroidal
envelope co-exist. The $N$-body simulations presented by \cite{ht}
show that flyby interactions impart angular momentum preferentially to
the outer parts of a galaxy (stars close to the galaxy's center have
much shorter orbital periods and respond adiabatically to the
perturbing forces during the interaction). The outer envelope's
acquired angular momentum is most likely not aligned with that of the
embedded disk and this results in a KDC-like signature in the velocity
profile. Multiple interactions with different impact parameters and
orbital angular momenta may produce the complex velocity profile
observed in FS373, with the rotation velocity changing sign. Moreover,
upon the first pericenter passage, the bar instability triggered in
the progenitor spiral galaxy funnels almost all the gas to the center
where it is consumed in a starburst. This process depletes the gas
within roughly 2~Gyr (\cite{ma}), effectively turning a gas-rich
late-type galaxy into a gas-poor dE.

\section{Conclusions} \label{con}

We have presented evidence for the discovery of two dwarf elliptical
galaxies with kinematically decoupled cores. The presence of a cold,
rotationally flattened subcomponent is supported by the photometry,
kinematics, and the dynamics of these galaxies. This is the first time
kinematically decoupled cores have been detected in dwarf elliptical
galaxies.

%Although the apparently slightly higher metallicity of the KDC hints
%at a gaseous merger followed by star-formation, as in giant
%ellipticals, 
KDCs in dwarf galaxies are not likely produced by mergers since the
gravitational field of a dwarf elliptical galaxy is not strong enough
for dynamical friction to sufficiently slow down another dwarf galaxy
at the relative velocities and impact parameters that are typical for
a group or cluster environment. In the field, relative velocities are
much lower but the galaxy density is prohibitively low for mergers to
occur. The harassment scenario is able to offer an alternative
explanation:~the angular momentum, transfered to the dwarf galaxy
during an encounter by the tidal forces, can result in the observed
peculiar kinematics. The fact that these two objects, the only ones in
a sample of 15 dEs with measured kinematics to host a KDC, are found
in a group environment (in which mergers are unlikely but interactions
are slow enough to transfer significant amounts of energy and angular
momentum) and not in a cluster environment (in which mergers are
unlikely and interactions are much too fast to transfer significant
amounts of angular momentum) agrees with this interpretation.

%Clearly, high-resolution imaging and spectroscopy in a wavelength
%region better suited for constraining the composition of stellar
%populations is required to pinpoint the origin of these KDCs.

\begin{acknowledgements}
Based on observations made at the European Southern Observatory, Chile
(ESO Large Programme Nr.~165.N-0115). SDR thanks P. Prugniel and
F. Simien for the hospitality and fruitful discussions during a stay
at the Observatoire de Lyon. WWZ acknowledges the support of the
Austrian Science Fund (project P14753). We like to thank the referee
J. A. L. Aguerri for his useful comments and suggested
improvements. This research has made use of the NASA/IPAC
Extragalactic Database (NED) which is operated by the Jet Propulsion
Laboratory, California Institute of Technology, under contract with
the National Aeronautics and Space Administration.
\end{acknowledgements}

\end{document}